\documentclass[universe,article,accept,pdftex,moreauthors]{mdpi} 
\firstpage{1} 
\pubvolume{8}
\issuenum{12}
\articlenumber{620}
\pubyear{2022}
\copyrightyear{2022}
\externaleditor{Academic Editor: {Arnab Dasgupta} %MDPI: Please provide academic editor.
}
\datereceived{24 October 2022} 
\dateaccepted{21 November 2022} 
\datepublished{25 November 2022} 
\hreflink{https://\linebreak doi.org/10.3390/universe8120620} % If needed use \linebreak
\usepackage{latexsym}
\usepackage{amsmath}
\usepackage{graphicx}
\usepackage{amsmath}
\usepackage{amssymb}
\usepackage{physics}
\usepackage{bm}
\def\UrlBreaks{\do\/\do-}
\Title{Description and Application of the Surfing Effect}
\TitleCitation{Description and Application of the Surfing Effect}

\Author{Michele Maiorano %Please carefully check the accuracy of names and affiliations. Changes will not be possible after proofreading.
$^{1,2,3,}$*\orcidA{}, Francesco De Paolis $^{1,2,3}$\orcidB{} and Achille A. Nucita $^{1,2,3}$}
\AuthorNames{Michele Maiorano, Francesco De Paolis and Achille Nucita}
\AuthorCitation{Maiorano, %Please carefully check the accuracy of names and affiliations. Changes will not be possible after proofreading.
M.; De Paolis, F.; Nucita, A.A.}
\address{
$^{1}$\quad Department of Mathematics and Physics ``Ennio De Giorgi'', University of Salento, Via Arnesano, \linebreak I-73100 %MDPI: Post code is identical for all affiliations. Please check and confirm. ------> AUTHOR: OK
 Lecce, Italy\\
$^{2}$ \quad Istituto Nazionale di Fisica Nucleare (INFN)%MDPI: Instituto was changed into Institute. Please confirm. Abbreviations were also revised. Please confirm. ------> AUTHOR: the names "Istituto Nazionale di Fisica Nucleare", "Istituto Nazionale di Astrofisica" are proper names.
, Sezione di Lecce, Via Arnesano, I-73100 Lecce, Italy\\
$^{3}$ \quad Istituto Nazionale di Astrofisica (INAF), Sezione di Lecce, Via Arnesano, I-73100 Lecce, Italy}

\corres{Correspondence: michele.maiorano@le.infn.it}
\abstract{The standard technique for very low-frequency gravitational wave detection is mainly based on searching for a specific spatial correlation in the variation of the times of arrival of the radio pulses emitted by millisecond pulsars with respect to a timing model. This spatial correlation, which in the case of the gravitational wave background must have the form described by the Hellings and Downs function, has not yet been observed. Therefore, despite the numerous hints of a common red noise in the timing residuals of many millisecond pulsars compatible with that expected for the gravitational wave background, its detection has not yet been achieved. By now, the reason is not completely clear and, from some recent works, the urgency to adopt new detection techniques, possibly complementary to the standard one, is emerging clearly. Of course, this demand also applies to the detection of continuous gravitational waves emitted by supermassive black hole binaries populating the Universe. In the latter case, important information could, in principle, emerge from the millisecond pulsars considered individually in a single-pulsar search of continuous GWs. In this context, the surfing effect can then be exploited, helping to select the best pulsars to carry out such analysis. This paper aims to clarify when the surfing effect occurs and describe it exhaustively. A possible application to the case of the supermassive black hole binary candidate PKS {2131--021}{} and millisecond pulsar J{2145--0750}{} is also analyzed.}
\begin{document}
%=================================================================
\section{Introduction}
\label{introduction}
\textls[-25]{{Since the first direct observation of gravitational waves (GWs), carried out by the interferometers of the LIGO/VIRGO collaboration \cite{abbott2016}, gravitational radiation has proved to be a precious messenger of information for astrophysics, allowing not only to confirm the predictions of general relativity \cite{einstein1916,depa2022} but also to study extraordinary phenomena, such as the merger of neutron stars \cite{abbott2017}, and to discover several black holes \cite{abbott2020}. However, gravitational astronomy is still in its infancy and to date, only the tip of the iceberg of the GW spectrum has been scratched. Additionally, low-frequency GWs are not accessible by ground-based interferometers, which are currently only sensitive to high-frequency GWs (in the range 10--$10^{4}$ Hz) emitted in the final stages of the coalescence of compact objects \cite{maggiore2008a}. To explore the low-frequency side of the GW spectrum (in the range $10^{-5}$--1 Hz), space-based gravitational interferometers, such as Laser Interferometer Space Antenna (LISA) \cite{lisa2017}, are required and, in addition, for very low-frequency GWs (in the range $10^{-10}$--$10^{-6}$ Hz), pulsar timing arrays (PTAs) must be used, which are currently the only instruments sensitive to them \cite{moore2015}.}

PTAs are arrays of extremely regular millisecond pulsars (MSPs), constantly monitored by radio telescopes to measure the variations in the times of arrival (ToAs) of the emitted radio pulses, commonly referred to as timing residuals (TRs). Since GWs can influence the radio pulse ToAs, their signature can be found by studying the TRs \cite{estabrook1975,sazhin1978,detweiler1979}, once cleaned from other effects that can affect them. At present, the main purpose of PTAs is to reveal the gravitational wave background (GWB) due to the superposition of GWs emitted by the supermassive black hole binary systems (SMBHBs) that, according to the current cosmological model, populate the Universe \cite{maggiore2008b}. The GWB should induce spatially correlated TRs in all the MSPs of PTAs \cite{mashhoon1982}. Since this correlation should take the form described by the Hellings and Downs function \cite{hellings1983}, the latter is considered the smoking gun for the GWB \cite{maiorano2021a}. Although from the data collected by the main PTA collaborations (the European Pulsar Timing Array (EPTA) \cite{desvignes2016}, the Indian Pulsar Timing Array (InPTA) \cite{joshi2018}, the North American Nanohertz Observatory for Gravitational Waves (NANOGrav) \cite{arzoumanian2018}, and the Parkes Pulsar Timing Array (PPTA) \cite{reardon2016}, which join their efforts as the International Pulsar Timing Array (IPTA) \cite{verbiest2016}) it has been possible to obtain strong constraints on both the frequency and the amplitude of the GWB, and despite the emergence from them of a common red noise compatible with that expected for the GWB, the smoking gun is still missing, and the first GW detection by PTAs has not yet been claimed. Furthermore, the first detection of continuous GW emitted by a SMBHB seems even more distant, as it may be subject to knowledge of the GWB.

\textls[-25]{This article is motivated by the exigency to look for additional instruments and tools for detecting very low-frequency GWs \cite{maiorano2021b}. This necessity is made more urgent by recent research works, which show the limits and criticalities of the investigation principles adopted up to now~\cite{zic2022}. For this purpose, the potential of the surfing effect for the detection of the continuous GWs emitted by a SMBHB has been analyzed. Although the surfing effect is present in some works on PTAs, the literature to date is lacking in articles dedicated to its discussion in the context of general relativity, and in which a possible concrete application is presented. For this reason, this paper is structured as follows: Section \ref{sec:1} is dedicated to a detailed description of the surfing effect, Section \ref{sec:2} describes an application of the surfing effect to the SMBHB candidate \linebreak PKS 2131--021, in Section \ref{sec:3} the signatures of the GWs possibly emitted by the SMBHB candidate \linebreak PKS 2131--021, have been searched in the averaged narrowband and wideband post-fit whitened TRs obtained by NANOGrav collaboration on the MSP J2145--0750, and finally, the main conclusions are presented Section \ref{sec:4}.}}

\section{An Overview on the Surfing Effect}
\label{sec:1}
The surfing effect occurs when the angle $\theta$ seen by an observer on Earth between the travel direction of the radio pulses emitted by a MSP and the travel direction of the GWs emitted by a single GW source is sufficiently small. In this case, the radio pulses behave as if they were surfing the GWs, producing an increased value of the pulsar TRs. That can be shown by considering the function $R\left(t,n_i\right)$ which describes the pulsar TRs induced by the GWs emitted by a single source \cite{estabrook1975, sazhin1978, detweiler1979}:
\begin{equation}
\label{residualsfunction}
R\left(t,n_i\right)=\sum_{A=+,\times}F^A\left[r^A_e\left(t,n_i\right) -r^A_p\left(t,n_i\right)\right]
\end{equation}
\textls[-25]{where $t$ is the time, $n_i$ is the versor oriented toward the source of the GWs\endnote{Sometimes, in the literature, the versor oriented along the travel direction of the GWs $\Omega_i$ is considered in place of $n_i$. \linebreak Since the two versors point toward opposite directions, the relation between them is $\Omega_i=-n_i$.}, $A$ is the GW polarization state index, which is $+$ in the case of the plus-polarization state, and $\times$ in the case of the cross-polarization state, and $F^A$ is the antenna pattern function, defined as\endnote{In this paper, the Einstein notation, which indicates the sum over repeated indices, has been adopted.}:}% MDPI: footnote is not allowed, we put these as Note before reference, please confirm. ------> AUTHOR: OK
\begin{equation}
\label{antennafunction}
F^A=\frac{1}{2}\frac{p^ip^je_{ij}^A}{(1-n_ip^i)}
\end{equation}

Here, $p^i$ is the versor oriented toward the MSP, and $e_{ij}^A$ is the GW polarization tensor:
\begin{equation}
\label{polarizationplus}
e_{ij}^+=\begin{pmatrix}
1& 0& 0 \\ 
0& -1& 0 \\ 
0& 0& 0 
\end{pmatrix}
\end{equation}
\begin{equation}
\label{polarizationscross}
e_{ij}^\times=\begin{pmatrix}
0& 1& 0 \\ 
1& 0& 0 \\ 
0& 0& 0 
\end{pmatrix}
\end{equation}

\textls[-35]{Moreover, $r^A_e\left(t,n_i\right)$ and $r^A_p\left(t,n_i\right)$ are the Earth and the pulsar terms defined respectively by\endnote{In this paper, the geometrical units c=G=1 have been adopted.}:}
\begin{equation}
\label{earthterm}
r^A_e\left(t,n_i\right)=\int^t_0\dd{t}h^A_e\left(\omega_e t\right)
\end{equation}
\begin{equation}
\label{pulsarterm}
r^A_p\left(t,n_i\right)=\int^t_0\dd{t}h^A_p\left(\omega_p t-\omega_p t_p\left(1-n_ip^i\right)\right)
\end{equation}
where $\omega_e$ and $\omega_p$ are the GW angular frequency, evaluated at the Earth and MSP positions, respectively, $t_p$ is the time distance from the MSP to the Earth, and $h^A_e\left(\omega_e t\right)$ and $h^A_p\left(\omega_p t-\omega_p t_p\left(1-n_ip^i\right)\right)$ are the perturbations on the flat space-time metric induced by the GWs, evaluated at the Earth and MSP positions, respectively\endnote{For an exhaustive derivation of Equation \eqref{residualsfunction} see ref. \cite{maggiore2008b}.}.

Equation \eqref{residualsfunction} is particularly useful to describe the pulsar TRs induced by the GWs emitted by a circularized SMBHBs in the continuous emission regime. In the literature, this assumption is often made in order to simplify the description of the principles on which PTAs are based. That is reasonable because, according to the most widely accepted models, SMBHBs form during the collision of galaxies by dynamic friction, which, as it is widely known, tends to circularize their orbits \cite{bonetti2020}.

\begin{figure}[H]
\hspace{-30pt}
\includegraphics[scale=0.65]{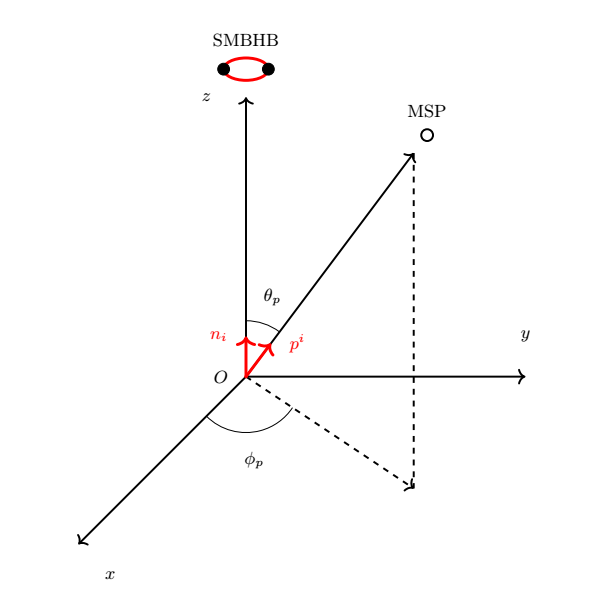}
\caption{Scheme of the adopted $Oxyz$ coordinate system, chosen so that the versor $n_i$ is aligned with the $z$-axis. The black circle pair indicates the SMBHB, pointed by the versor $n_i$, while the white circle indicates the MSP, pointed by the versor $p_i$. In the figure, the distances of the SMBHB and the MSP from the origin $O$, where the observer is placed, are not in scale.}
\label{fig:reference}
\end{figure}

In this case, it is convenient to adopt a coordinate system $Oxyz$ with its origin on the Earth\endnote{Such a choice would induce spurious TRs, with a period of one year, due to the motion of the Earth relative to the Solar System Barycenter. Although accounting for that is crucial from the experimental point of view and is done during the data processing phase, this issue can be safely ignored in this theoretical context.}, so that:
\begin{equation}
\label{nversor}
n_i\equiv\left(0,0,1\right)
\end{equation}
\begin{equation}
\label{pversor}
p^i\equiv\left(\sin{\theta_p}\cos{\phi_p},\sin{\theta_p}\sin{\phi_p},\cos{\theta_p}\right)
\end{equation}
\textls[-35]{where $\theta_p$ is the angle between the versor $p^i$ and the positive $z$-axis, and $\phi_p$ is the angle between the $x$-axis and the projection on the $xy$-plane of the versor $p^i$ (see Figure \ref{fig:reference}). With that choice, the product $n_ip^i$, which appears in Equations \eqref{antennafunction}, \eqref{earthterm}, and \eqref{pulsarterm}, is simply $\cos\theta_p$.}

The antenna pattern function in Equation \eqref{antennafunction} can be rewritten in this coordinate system, taking into account the Equations \eqref{polarizationplus} and \eqref{polarizationscross}, for both the polarization states:
\begin{equation}
\label{antennaplus}
F^+=\frac{1}{2}\frac{\sin^2\theta_p}{1-\cos\theta_p}\cos(2\phi_p)
\end{equation}
\begin{equation}
\label{antennacross}
 F^\times=\frac{1}{2}\frac{\sin^2\theta_p}{1-\cos\theta_p}\sin(2\phi_p)
\end{equation}

Moreover, the perturbations on the flat space-time metric induced by the GWs evaluated at the Earth and MSP positions are \cite{perrodin2018}:
\begin{equation}
\label{hearthterm}
h^A_e\left(\omega_e t\right)=\mathcal{F}^A\mathcal{A}\sin{\left(\omega_e t+\alpha^A\right)}
\end{equation}
\begin{equation}
\label{hpulsarterm}
h^A_p\left(\omega_p t-\omega_p t_p\left(1-\cos\theta_p\right)\right)=\mathcal{F}^A\mathcal{A}\sin{\left(\omega_p t-\omega_p t_p\left(1-\cos\theta_p\right)+\alpha^A\right)}
\end{equation}
where $\mathcal{F}^A$ is a factor that accounts for the orbital inclination angle $\iota$ of the SMBHB \cite{sesana2010}:
\begin{equation}
\label{inclinationplus}
\mathcal{F}^+=1+\cos^2{\iota}
\end{equation}
\begin{equation}
\label{inclinationcross}
\mathcal{F}^\times=-2\cos{\iota}
\end{equation}
and $\mathcal{A}$ is the perturbation amplitude, given by:
\begin{equation}
\label{amplitude}
\mathcal{A}=2\frac{\mathcal{M}^{5/3}}{D}\left(\frac{\omega}{2}\right)^{2/3}
\end{equation}
where $\mathcal{M}$ is the chirp mass of the SMBHB, $D$ is the luminosity distance, and $\omega$ is the angular frequency of GWs, and $\alpha^A$ is the initial phase, with $\alpha^+=0$ and $\alpha^\times=\pi/2$. Note that all the quantities in Equation \eqref{amplitude} are red-shifted \cite{perrodin2018}.

The difference between the Earth term, in Equation \eqref{earthterm}, and the pulsar term, \linebreak in Equation \eqref{pulsarterm}, can be determined for both the polarization states by taking into account Equations \eqref{hearthterm}--\eqref{amplitude}, obtaining:
\begin{equation}
\label{differenceplus}
\begin{split}
&r^+_e\left(t,n_i\right) -r^+_p\left(t,n_i\right)=-\frac{\mathcal{A}}{\omega_e}\left(1+\cos^2{\iota}\right)\left[\cos{\left(\omega_e t\right)}-1\right]+\\
&+\frac{\mathcal{A}}{\omega_p}\left(1+\cos^2{\iota}\right)\left[\cos{\left(\omega_p t-\omega_p t_p\left(1-\cos\theta_p\right)\right)}-\cos{\left(-\omega_p t_p\left(1-\cos\theta_p\right)\right)}\right]
\end{split}
\end{equation}
\begin{equation}
\label{differencecross}
\begin{split}
&r^\times_e\left(t,n_i\right) -r^\times_p\left(t,n_i\right)=-\frac{\mathcal{A}}{\omega_e}\left(-2\cos{\iota}\right)\left[-\sin{\left(\omega_e t\right)}\right]+\\
&+\frac{\mathcal{A}}{\omega_p}\left(-2\cos{\iota}\right)\left[-\sin{\left(\omega_p t-\omega_p t_p\left(1-\cos\theta_p\right)\right)}+\sin{\left(-\omega_p t_p\left(1-\cos\theta_p\right)\right)}\right]
\end{split}
\end{equation}

Then, the full expression of the function $R\left(t,n_i\right)$ is found by substituting the Equations \eqref{antennaplus}, \eqref{antennacross}, \eqref{differenceplus} and \eqref{differencecross} in Equation \eqref{residualsfunction}. The result obtained above can be simplified by introducing some additional working hypotheses. First of all, the SMBHB can be assumed to be face-on, which means that $\iota=\pi/2$. In this case, as can be seen from Equations \eqref{inclinationplus} and \eqref{inclinationcross}, $\mathcal{F}^+=1$ and $\mathcal{F}^\times=0$. Secondly, since only one MSP is under consideration, the angle $\theta_p$ coincides with the angle $\theta$, so $\theta_p=\theta$, the coordinate system is rotated to have $\phi_p=0$, and $t_p$ is replaced by $T$ to simplify the notation. In this case, as can be seen by Equations $\eqref{antennaplus}$ and \eqref{antennacross}, $F^+=\sin^2\theta\left[2\left(1-\cos\theta\right)\right]^{-1}$ and %MDPI: Is the bold format of the previous equation necessary? Please confirm. Same for all bold italic formatted text from within the manuscript. ------> AUTHOR: it is not necessary, I used bf to highlight the previous changes after the first submission. Same for others.
$F^\times=0$. Thirdly, the SMBHB can be assumed to be very far from its coalescence, as most of the SMBHBs observable with PTAs should be. Under this assumption, the variation of the angular frequency of the GWs~\cite{shapiro1983} during the interval of time $\Delta t=T\left(1-\cos\theta\right)$, which denotes the time delay between the Earth epoch and the pulsar epoch \cite{jenet2004}, can be neglected, so $\omega_e\approx\omega_p$ and the angular frequency of GWs can be denoted by just $\omega$. Using all these assumptions, one gets:
\begin{equation}
\label{compactresiduals}
\begin{split}
&R\left(t,\theta\right)=-\frac{1}{2}\frac{\sin^2\theta}{1-\cos\theta}\frac{\mathcal{A}}{\omega}\left[\cos{\left(\omega t\right)}-1\right]+\\&-\frac{1}{2}\frac{\sin^2\theta}{1-\cos\theta}\frac{\mathcal{A}}{\omega}\left[-\cos{\left(\omega t-\omega T\left(1-\cos\theta\right)\right)}+\cos{\left(-\omega T\left(1-\cos\theta\right)\right)}\right]
\end{split}
\end{equation}
where the dependence on $n_i$ has been replaced by the dependence on $\theta$. As can be seen from Equation \eqref{compactresiduals}, the way the function $R\left(t,\theta\right)$ depends on the angle $\theta$ makes its use for the detection of GWs rather problematic. For example, it is interesting to note that $R\left(t,\theta\right)$ has a discontinuity for $\theta=0$, which can be removed by placing $R\left(t,0\right)\equiv\lim_{\theta\to 0} R\left(t,\theta\right)=0$. Moreover, for values of $\omega$ and $T$ of the same order of magnitude of the average angular frequency of the GWs emitted by SMBHB observable by PTAs and of the average time distance of the MSPs included in PTAs, respectively, the function $R\left(t,\theta\right)$ oscillates extremely rapidly with $\theta$.
 
Since the function $R(t,\theta)$ carries all the information about the GWs emitted by the SMBHB, it is crucial to have a clear understanding of its behavior. The easiest way to do this is by considering, instead of $R\left(t,\theta\right)$, the function $S(\zeta,\eta)$, which, using the substitutions:
\begin{equation}
\label{sostitutionzeta}
\zeta\equiv\chi\frac{t}{T}
\end{equation}
\begin{equation}
\label{sostitutioneta}
\eta\equiv\chi\left(1-\cos\theta\right)
\end{equation}
where $\chi\equiv\omega T$, $\zeta\in\left[0,\infty\right[$ for $t\in\left[0,\infty\right[$ and $\eta\in\left[0,2\chi\right]$ for $\theta\in\left[0,\pi\right]$, can be defined by:
\begin{equation}
\label{sfunctiondef}
S\left(\zeta,\eta\right)\equiv2\frac{\chi}{\mathcal{A}T}\left|R\left(\zeta,\eta\right)\right|
\end{equation}

The function defined in Equation \eqref{sfunctiondef} can be written more explicitly as:
\begin{equation}
\label{sfunction}
S\left(\zeta,\eta\right)=\left(2-\frac{\eta}{\chi}\right)\left|\cos\zeta-1-\cos\left(\zeta-\eta\right)+\cos(-\eta)\right|
\end{equation}
\textls[-25]{where the first factor in Equation \eqref{sfunction} has taken out the absolute value since it is always positive in the defined domain. The TRs induced by GWs are enhanced for the values which maximize the function $S\left(\zeta,\eta\right)$ in Equation \eqref{sfunction}. The function $S\left(\zeta,\eta\right)$ is a periodic function characterized by an amplitude that depends only on $\eta$ through the first linear factor:}
\begin{equation}
\label{samplitude}
a\left(\eta\right)\equiv 2-\frac{\eta}{\chi}
\end{equation}
and its shape depends on both the variables $\zeta$ and $\eta$ through the second oscillating factor:
\begin{equation}
\label{soscillating}
f\left(\zeta,\eta\right)\equiv\left|\cos\zeta-1-\cos\left(\zeta-\eta\right)+\cos(-\eta)\right|
\end{equation}
so that:
\begin{equation}
\label{shortsfunction}
S\left(\zeta,\eta\right)=a\left(\eta\right)f\left(\zeta,\eta\right)
\end{equation}

Equation \eqref{shortsfunction} allows to find the global maximum of the function $S\left(\zeta,\eta\right)$ by studying the maxima of the functions $a\left(\eta\right)$ and $f\left(\zeta,\eta\right)$. The function $a\left(\eta\right)$ is maximized when: 
\begin{equation}
\label{firstfactor}
a\left(\eta\right)=2\text{ for }\eta=0
\end{equation}
that occurs when:
\begin{equation}
\label{surfingeffect}
\theta=0
\end{equation}
 while the function $f\left(\zeta,\eta\right)$ is maximized when: 
\begin{equation}
\label{secondfactor}
f\left(\zeta,\eta\right)=4\text{ for }\left(\zeta,\eta\right)=\left(\pi+2\pi n_\zeta,\pi+2\pi n_\eta\right)
\end{equation}
that occurs when:
\begin{equation}
\label{timeposition}
t=\frac{\pi T}{\chi}\left(1+2n_\zeta\right)\equiv t_{n_\zeta}
\end{equation}
\begin{equation}
\label{angularposition}
\theta=\arccos\left(1-\frac{\pi}{\chi}\left(1+2n_\eta\right)\right)\equiv\theta_{n_\eta}
\end{equation}

Therefore, the results in Equations \eqref{surfingeffect} and \eqref{angularposition} imply that, for each value of $t$, the global maximum of the function $R\left(\zeta,\eta\right)$ can be found placing
$n_{\eta}\equiv 0$. Then, since for $n_{\eta}= 0$ results $\theta_0\ll 1$, this property is referred to as surfing effect, and $\theta_0=\arccos\left(1-\pi/\chi\right)$ is the surfing effect angle.

\section{An Application of the Surfing Effect to the Case of the Supermassive Black Hole Binary Candidate PKS 2131--021}
\label{sec:2}
The existence of SMBHBs is not only predicted by most of the hierarchical structure formation models \cite{sesana2013} but also supported by observational evidence. In some rare cases, the inferred orbital parameters suggest that, if the data analysis is correct, the SMBHB should emit GWs observable with PTAs. In these cases, as has already been conducted with the SMBHB candidate in the Seyfert galaxy $3$C $66$B, the data collected by PTAs can be used to rule out the GW emission \cite{jenet2004,depa2004,arzoumanian2020a}.

An interesting recent study on the blazar PKS 2131--021 opens the possibility for an application of the surfing effect \cite{oneill2022}. In fact, according to the data analysis of the blazar PKS 2131--021 radio emission, it can be identified as an SMBHB candidate located at a redshift $z=1.285$ \cite{wu2007}, with an observed orbital period of $1760.4^{+5.3}_{-5.3}$ d and a non-red-shifted chirp mass $\lesssim 5.4 \times 10^9 M_\odot$ (see ref. \cite{oneill2022}). Adopting the most recent values for cosmological constants determined by the Planck collaboration \cite{planck2020}, the luminosity distance of the SMBHB candidate is $D=9.2$ Gpc, while the angular frequency and the amplitude of the expected GW emission are $\omega=8.262^{+0.025}_{-0.025}\times 10^{-8}$ Hz and $\mathcal{A}\lesssim 2.4\times 10^{-15}$, respectively.

An optimal strategy to confirm these results, by taking advantage of the surfing effect, is to perform a single-pulsar search of continuous GWs focusing on the MSP with the smallest angular distance from the SMBHB candidate PKS 2131--021 among the ones currently included in PTAs. For this reason, the best MSP is the MSP J2145--0750, which lies at the angle $\theta=0.1156$, corresponding to about $6.6234^{\circ}$, from the SMBHB candidate PKS 2131--021 \cite{gaia,deller2019}. The MSP J2145--0750 is a well-studied MSP, lying in a binary system with a white dwarf, at a distance of $0.62^{+0.00}_{-0.02}$ kpc \cite{deller2019} from the Earth and timed for about $12.5$~years by the NANOGrav collaboration \cite{alam2021a, alam2021b}. The choice of the MSP J2145--0750 is also convenient because it allows neglecting the GW angular frequency variation between the Earth and the pulsar terms due to the SMBHB PKS 2131--021 orbital evolution. In fact, by considering the equation describing the GW angular frequency variation \cite{shapiro1983}:
\begin{equation}
\label{frequencyevolution}
\dot{\omega}=\frac{12}{5}2^{1/3}\mathcal{M}^{5/3}\omega^{11/3}\left(1+z\right)
\end{equation}
and substituting in the integral of Equation \eqref{frequencyevolution} the time delay between the Earth epoch and the pulsar epoch, which is $\Delta t = 13.5\pm 0.4$ yr, the GW angular frequency variation between the Earth and the pulsar terms turns out to be $\Delta\omega\lesssim0.042\times 10^{-8}$ Hz.

\begin{figure}[H]

\includegraphics[scale=0.85]{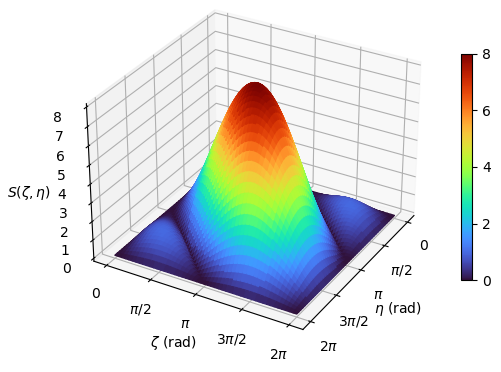}\hfill
\caption{Plot of the $S\left(\zeta,\eta\right)$ function relative to the SMBHB candidate PKS 2131--021 and the MSP J2145--0750. The axes on the plane indicate the values assumed by the $\zeta$ and $\eta$ variables defined in Equations \eqref{sostitutionzeta} and \eqref{sostitutioneta}, respectively, expressed in radians. The vertical axis and the color, described by the color bar, indicate the values assumed by $S\left(\zeta,\eta\right)$, defined in Equation \eqref{sfunctiondef}, dimensionless.}
\label{fig:sfunction}
\end{figure}

\begin{figure}[H]
\hspace{-5pt}
\includegraphics[scale=0.53]{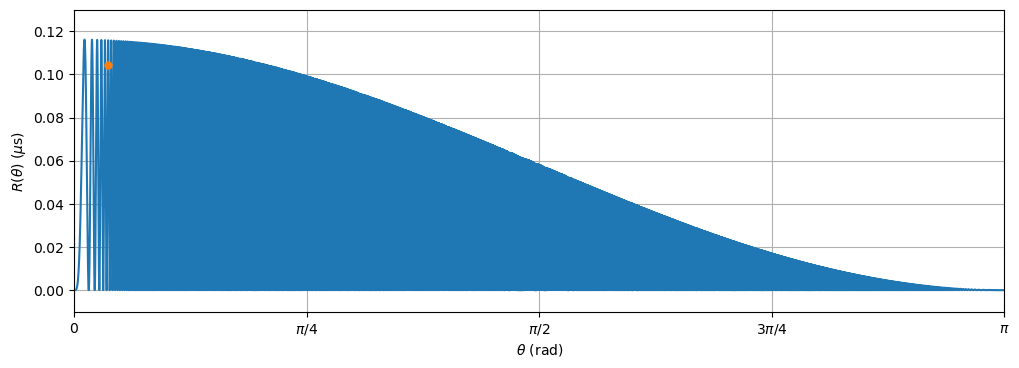}\hfill
\caption{{Plot} 
 of the $R\left(\theta\right)$ function relative to the SMBHB candidate PKS 2131--021 and the MSP J2145--0750. The horizontal axis indicates the values assumed by the $\theta$ variable, expressed in radians. The vertical axis indicates the values assumed by $R\left(\theta\right)$, expressed in nanoseconds. The orange dot indicates the value assumed by the $R\left(\theta\right)$ function when $\theta$ coincides with the actual angle between the SMBHB candidate PKS 2131--021 and the MSP J2145--0750, which is $\theta=0.1156$.}
\label{fig:seangle}
\end{figure}

\begin{figure}[H]
\hspace{-5pt}
\includegraphics[scale=0.53]{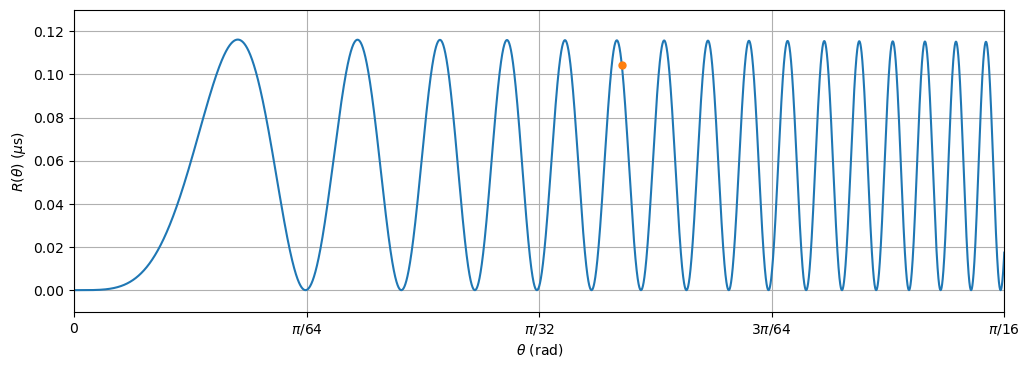}\hfill
\caption{Zoomed plot of the $R\left(\theta\right)$ function relative to the SMBHB candidate PKS 2131--021 and the MSP J2145--0750. The horizontal axis indicates the values assumed by the $\theta$ variable, expressed in radians. The vertical axis indicates the values assumed by the $R\left(\theta\right)$ function, expressed in nanoseconds. The orange dot indicates the value assumed by $R\left(\theta\right)$ when $\theta$ coincides with the actual angle between the SMBHB candidate PKS 2131--021 and the MSP J2145--0750, which is $\theta=0.1156$.}
\label{fig:seanglezoom}
\end{figure}

\begin{figure}[H]
\hspace{-5pt}
\includegraphics[scale=0.53]{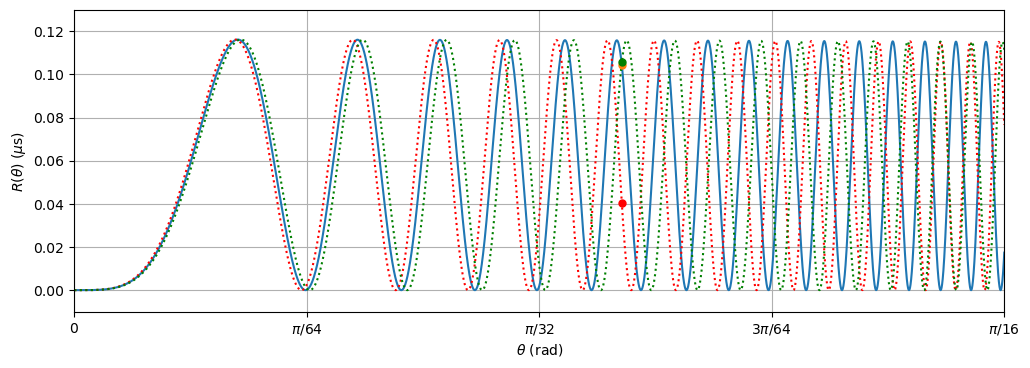}\hfill
\caption{Comparative zoomed plot of the $R\left(\theta\right)$ function relative to the SMBHB candidate PKS 2131--021 and the MSP J2145--0750. The horizontal axis indicates the values assumed by the $\theta$ variable, expressed in radians. The vertical axis indicates the values assumed by the $R\left(\theta\right)$ function, expressed in nanoseconds. The green and red dotted lines indicate the values assumed by the $R\left(\theta\right)$ function for $\chi=5274-186$ and $\chi=5274+58$, respectively. The orange dot indicates the value assumed by $R\left(\theta\right)$ when $\theta$ coincides with the actual angle between the SMBHB candidate PKS 2131--021 and the MSP J2145--0750, which is $\theta=0.1156$. The green and red dots indicate the values assumed by $R\left(\theta\right)$ when $\theta$ coincides with the actual angle between the SMBHB candidate PKS 2131--021 and the MSP J2145--0750, which is $\theta=0.1156$, for $\chi=5274-186$ and $\chi=5274+58$, respectively.}
\label{fig:seanglezoomerror}
\end{figure}

With all these data in hand, Equation \eqref{angularposition} can be used to determine the function $S\left(\zeta,\eta\right)$ and then to find out the surfing effect angle (see Figure \ref{fig:sfunction}).
Once $t$ has been chosen to satisfy Equation \eqref{timeposition}, by setting, for example, $n_\zeta\equiv0$, it is possible to define the function $R\left(\theta\right)\equiv R\left(t_0,\theta\right)$, useful for verifying whether the angular separation $\theta$ between the MSP and the SMBHB candidate is such as to maximize the TRs. In this case, $\omega=8.262^{+0.025}_{-0.025}\times 10^{-8}$ Hz and $T=2024^{+16}_{-65}$ yr, therefore $\chi=5274^{+58}_{- 186}$, and at the angle $\theta_0=0.0345$, corresponding to about $1.9778^{\circ}$, one has $R\left(\theta_0\right)\lesssim 0.116$ $\upmu$s. Even if the actual angle between the SMBHB candidate PKS 2131--021 and the MSP J2145--0750, which is $\theta=0.1156$, is larger than $\theta_0$, it can be found that this is very close to $\theta_5=0.1145$, which identifies the sixth peak of the function $R\left(\theta\right)$. In fact, for $\theta=0.1156$ results $R\left(\theta\right)\lesssim0.104$ $\upmu$s (see Figures \ref{fig:seangle} and \ref{fig:seanglezoom}).
We emphasize that this is a lucky coincidence and allows one to say that if the chirp mass of the SMBHB candidate PKS 2131--021 coincides with the estimated upper limit, the TRs induced by the GWs emitted by the SMBHB candidate PKS 2131--021 on the MSP J2145--0750 might be observable even at the current PTA sensitivity level due to the surfing effect. In addition, as can be easily deduced from the Figures \ref{fig:seangle} and \ref{fig:seanglezoom}, if $\theta$, whether small or large, is in a node of the function $R\left(\theta\right)$, where it vanishes, in the TRs of the considered MSP there will be no trace of the influence of GWs. Eventually, it is necessary to stress the fact that the larger is $\theta$, the higher the precision required for $\chi$. In fact, as it is shown in \mbox{Figure \ref{fig:seanglezoomerror}}, for the same value of $\theta$ the function $R\left(\theta\right)$ can assume sensibly different values within the uncertainty on $\chi$. However, if $\theta$ is in the first peak, the role played by the error on $\chi$ is marginal.

\section{A Look to NANOGrav Data}
\label{sec:3}
\begin{figure}[H]
\hspace{-5pt}
\includegraphics[scale=0.53]{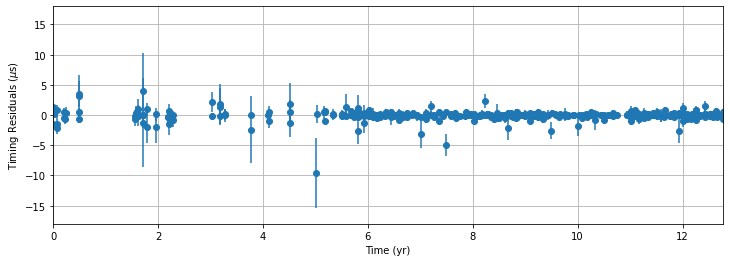}\hfill
\caption{Plot of the averaged narrowband post-fit whitened TRs of the MSP J2145--0750, which are from the $12.5$ yr Data Set published by the NANOGrav collaboration. The horizontal axis indicates the time, expressed in years. The vertical axis indicates the averaged narrowband post-fit whitened TRs, expressed in~nanoseconds.}
\label{fig:narrow1}
\end{figure}

\begin{figure}[H]

\includegraphics[scale=0.53]{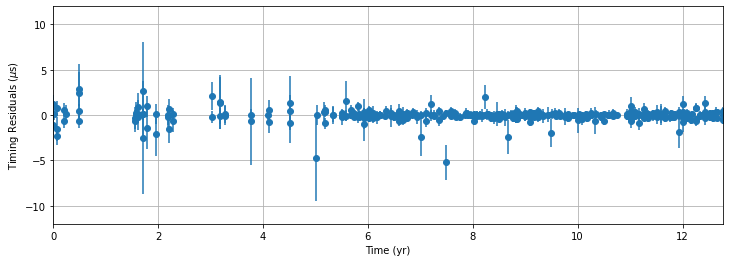}\hfill
\caption{Plot of the wideband post-fit whitened TRs of the MSP J2145--0750, which are from the $12.5$ yr Data Set published by the NANOGrav collaboration. The horizontal axis indicates the time, expressed in years. The vertical axis indicates the wideband post-fit whitened TRs, expressed in~nanoseconds.}
\label{fig:wide1}
\end{figure}

The NANOGrav collaboration has recently published the $12.5$ yr Data Set, %MDPI: please confirm this format ------> AUTHOR: the official name is "NANOGrav 12.5 yr Data Set"
which is publicly available (see ref. \cite{nanoweb}), making it possible to examine the TRs of the MSP J2145--0750. In particular, the datasets useful for searching GWs are the averaged narrowband \cite{alam2021a} and wideband \cite{alam2021b} post-fit whitened TRs (see Figures \ref{fig:narrow1} and \ref{fig:wide1}) since they are characterized by a low weighted root mean squared value $\sigma_w$, which has been evaluated by using the following definition:
\begin{equation}
\label{wrms}
\sigma_w = \sqrt{\frac{\sum_{i}^n{\frac{\left(x_i-\Bar{x}\right)^2}{\sigma_i^2}}}{\sum_{i}^n(\frac{1}{\sigma_i^2})}}
\end{equation}
which is valid for any set of quantities $x_1, x_2, \ldots, x_n$ characterized by the errors $\sigma_1, \sigma_2, \ldots, \sigma_n $ and by an average $\Bar{x}$. In this case, using Equation \eqref{wrms} it has been found $\sigma_w=0.333$ $\upmu$s for narrowband, and $\sigma_w=0.276$ $\upmu$s for wideband.
\vspace{-6pt}
In Figures \ref{fig:narrow1} and \ref{fig:wide1}, the origin of time is at Modified Julian Date $53267$, corresponding to 19 September $2004$, from which the observation started. Equation \eqref{compactresiduals} implies that the TRs induced by the GWs emitted by a SMBHB are periodic, with an angular frequency that is the same as the angular frequency $\omega$ of the GWs. This periodic signature can be searched in TRs using periodic analysis algorithms, such as the Lomb-Scargle (LS) algorithm \cite{lomb1976,scargle1982}.

\begin{figure}[H]

\includegraphics[scale=0.53]{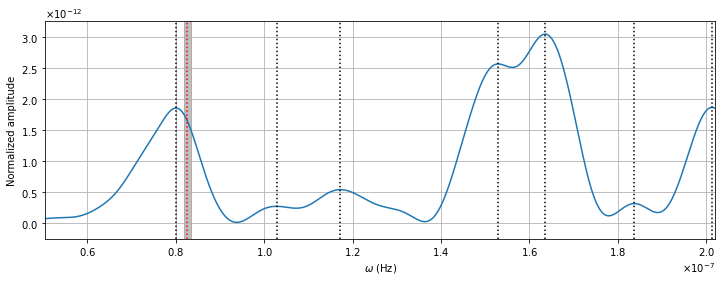}\hfill
\caption{Plot %MDPI: Please change the terms into scientific notations in the figure, e.g., “8 × 10³”, not “8E3”. ------> AUTHOR: OK
of the LS periodogram of the averaged narrowband post-fit whitened TRs of the MSP J2145--0750. The black-dotted lines indicate the angular frequencies found by the LS algorithm. The red-dotted line indicates the angular frequency $\omega=8.262^{+0.025}_{-0.025}\times 10^{-8}$ expected for the GWs possibly emitted by the SMBHB candidate PKS 2131--021. The gray shaded area indicates uncertainty on $\omega$ at the $3\sigma$ confidence level. The horizontal axis indicates the GW angular frequency, expressed in Hertz. The vertical axis indicates the normalized amplitude, dimensionless.}
\label{fig:narrow2}
\end{figure}

\begin{figure}[H]

\includegraphics[scale=0.53]{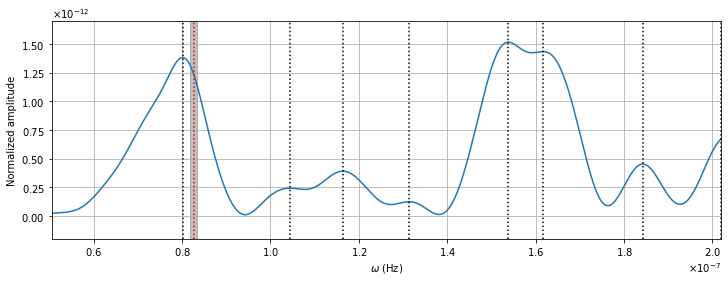}\hfill
\caption{Plot %MDPI: Please change the terms into scientific notations in the figure, e.g., “8 × 10³”, not “8E3”.  ------> AUTHOR: OK
of the LS periodogram of the wideband post-fit whitened TRs of the MSP J2145--0750. The black-dotted lines indicate the angular frequencies found by the LS algorithm. The red-dotted line indicates the angular frequency $\omega=8.262^{+0.025}_{-0.025}\times 10^{-8}$ expected for the GWs possibly emitted by the SMBHB candidate PKS 2131--021. The gray shaded area indicates uncertainty on $\omega$ at the $3\sigma$ confidence level. The horizontal axis indicates the GW angular frequency, expressed in Hertz. The vertical axis indicates the normalized amplitude, dimensionless.}
\label{fig:wide2}
\end{figure}

The LS periodograms of the TRs plotted in Figures \ref{fig:narrow1} and \ref{fig:wide1} have been obtained and plotted in Figures \ref{fig:narrow2} and \ref{fig:wide2}, respectively.
\vspace{-12pt}
In both cases (see Figures \ref{fig:narrow2} and \ref{fig:wide2}), among the periodicities with the highest normalized amplitude, the LS algorithm finds one associated with an angular frequency close to \linebreak $\omega=8.262^{+0.025}_{-0.025}\times 10^{-8}$ Hz, expected for GWs possibly emitted by the SMBHB candidate PKS 2131--021. Specifically, it finds $\omega=8.006\times 10^{-8}$ Hz in the case of the narrowband dataset and $\omega=8.0139\times 10^{-8}$ Hz in the case of the wideband dataset. At this point, a shuffling test was conducted to verify that these values are not artifacts and emerge from true periodicities in the datasets. To this aim, several datasets were created by shuffling the starting datasets, and, for each of the shuffled datasets, the LS periodogram was obtained. After doing so, the periodogram associated with the starting dataset was compared with each periodogram associated with the shuffled datasets to establish which is characterized by the largest normalized amplitude at the frequency value under consideration. \linebreak This procedure was iterated 10,000 times, with the result that the former normalized amplitude was larger with respect to the latter the $\simeq$87\% of cases for narrowband and the $\simeq$93\% of cases for wideband.

\begin{figure}[H]

\includegraphics[scale=0.53]{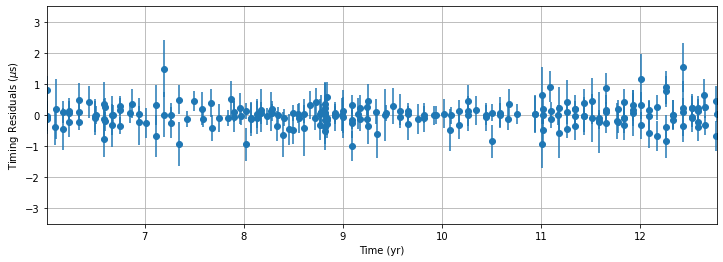}\hfill
\caption{Plot of the averaged narrowband post-fit whitened TRs of the MSP J2145--0750, filtered by removing the TRs associated with an uncertainty larger than $3\sigma_w$ and the TRs before the sixth year of observation. The horizontal axis indicates the time, expressed in years. The vertical axis indicates the averaged narrowband post-fit whitened TRs, expressed in nanoseconds.}
\label{fig:narrow3}
\end{figure}

\begin{figure}[H]

\includegraphics[scale=0.53]{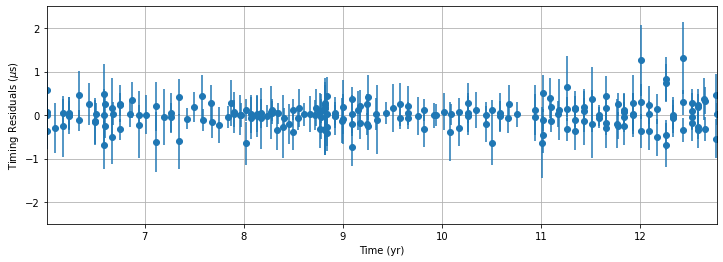}\hfill
\caption{Plot of the wideband post-fit whitened TRs of the MSP J2145--0750, filtered by removing the TRs associated with an uncertainty larger than $3\sigma_w$ and the TRs before the sixth year of observation. The horizontal axis indicates the time, expressed in years. The vertical axis indicates the averaged narrowband post-fit whitened TRs, expressed in nanoseconds.}
\label{fig:wide3}
\end{figure}

\begin{figure}[H]

\includegraphics[scale=0.53]{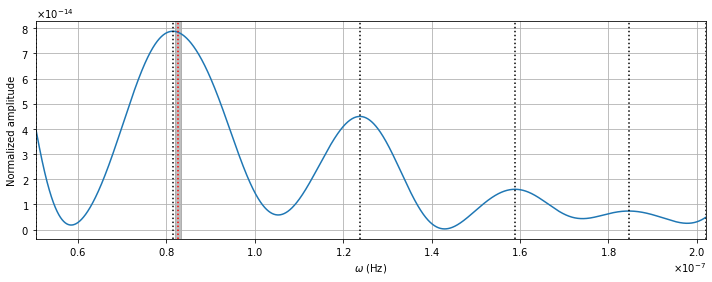}\hfill
\caption{Plot %MDPI: Please change the terms into scientific notations in the figure, e.g., “8 × 10³”, not “8E3”. ------> AUTHOR: OK
of the LS periodogram of the averaged narrowband post-fit whitened TRs of the MSP J2145--0750, filtered by removing the TRs associated with an uncertainty larger than $3\sigma_w$ and the TRs before the sixth year of observation. The black-dotted lines indicate the angular frequencies found by the LS algorithm. The red-dotted line indicates the angular frequency $\omega=8.262^{+0.025}_{-0.025}\times 10^{-8}$ expected for the GWs possibly emitted by the SMBHB candidate PKS 2131--021. The gray shaded area indicates uncertainty on $\omega$ at the $3\sigma$ confidence level. The horizontal axis indicates the GW angular frequency, expressed in Hertz. The vertical axis indicates the normalized amplitude, dimensionless.}
\label{fig:narrow4}
\end{figure}

\begin{figure}[H]
\includegraphics[scale=0.53]{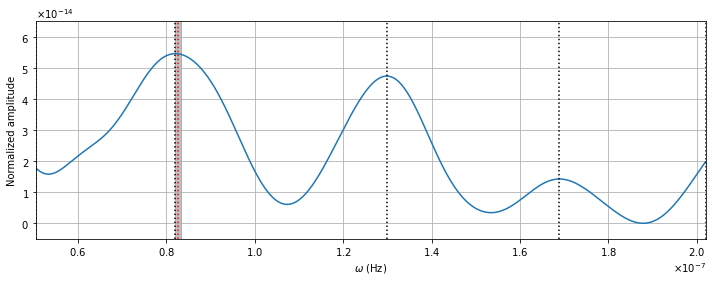}\hfill
\caption{Plot %MDPI: Please change the terms into scientific notations in the figure, e.g., “8 × 10³”, not “8E3”. ------> AUTHOR: OK
of the LS periodogram of the wideband post-fit whitened TRs of the MSP J2145--0750, filtered by removing the TRs associated with an uncertainty larger than $3\sigma_w$ and the TRs before the sixth year of observation. The black-dotted lines indicate the angular frequencies found by the LS algorithm. The red-dotted line indicates the angular frequency $\omega=8.262^{+0.025}_{-0.025}\times 10^{-8}$ expected for the GWs possibly emitted by the SMBHB candidate PKS 2131--021. The gray shaded area indicates uncertainty on $\omega$ at the $3\sigma$ confidence level. The horizontal axis indicates the GW angular frequency, expressed in Hertz. The vertical axis indicates the normalized amplitude, dimensionless.}
\label{fig:wide4}
\end{figure}

Although it can be read as an encouraging result, it is important to stress that the two periodograms have been obtained using the complete datasets, provided by the NANOGrav collaboration. However, from Figures \ref{fig:narrow1} and \ref{fig:wide1}, it can be noted that, in the first observation years, the cadence of TRs was sporadic and irregular, and the error bands were significantly large than in the last observation years. Therefore, it may be convenient to reanalyze the datasets after applying some filters. Based on the previous considerations, one possibility is to remove both the TRs falling in the first six observation years and those associated with an uncertainty larger than $3\sigma_w$ (see Figures \ref{fig:narrow3} and \ref{fig:wide3}). Applying this filter also assures that, as shown in ref. \cite{oneill2022}, the dataset is limited to a time interval from 2010 onwards when the SMBHB candidate should have been in the continuous emission regime.
\vspace{-12pt}
The LS periodograms of the TRs plotted in Figures \ref{fig:narrow3} and \ref{fig:wide3} have been obtained and plotted in Figures \ref{fig:narrow4} and \ref{fig:wide4}, respectively.
\vspace{-12pt}
In both cases (see Figures \ref{fig:narrow4} and \ref{fig:wide4}), the LS algorithm allows finding the periodicity with the highest normalized amplitude associated with an angular frequency even closer to $\omega=8.262^{+0.025}_{-0.025}\times 10^{-8}$ Hz. Specifically, it is found $\omega=8.147\times 10^{-8}$ Hz in the case of the narrowband dataset, and $\omega=8.205\times 10^{-8}$ Hz in the case of the wideband dataset. Even if these values are so close to the angular frequency of the GW emission expected for the SMBHB candidate PKS 2131--021, this is not enough for claiming its detection. Further, more sophisticated analyses must be conducted in order of ensuring the solidity of this result and the nature of this periodicity.

\section{Conclusions}
\label{sec:4}
In this article, a possible investigation method to search for continuous GWs, based on the surfing effect, has been proposed. The surfing effect, described in Section \ref{sec:1}, is usually ignored in most of the studies on the GWB since it is not relevant when the entire array of MSPs is used for its detection \cite{hellings1983,lentati2015,arzoumanian2020,goncharov2021}. However, when performing a single-pulsar search of GWs emitted by a circularized SMBHBs in the continuous emission regime, it has to be taken into consideration. In fact, in this case, as shown in Sections \ref{sec:1} and \ref{sec:2}, the global maximum of the function $R\left(t,\theta\right)$, which describes the TRs induced by GWs, corresponds, at any time, to an angular separation between the SMBHB and the MSPs almost null. However, it is important to keep in mind that, at any time, the function $R\left(t,\theta\right)$ is characterized by very rapid angular oscillations, the more frequent the greater the $\chi$ factor, as shown in Figures \ref{fig:seangle} and \ref{fig:seanglezoom}. Therefore, the function $R\left(\theta\right)$ represents a sort of map, which can be used to verify if a MSP is more or less suitable for performing a single-pulsar search of continuous GWs, so MSPs that fall on one of the peaks will be preferred, and those that fall into one of the nodes will be discarded.

As an example, in Sections \ref{sec:2} and \ref{sec:3}, the case of the SMBHB candidate PKS 2131--021 was considered. This system, based on the orbital parameters determined by the radio analysis of its emission \cite{oneill2022}, should be an emitter of GWs satisfying the hypotheses made in Section \ref{sec:1} for surfing effect. By a lucky coincidence, as shown in \mbox{Figures \ref{fig:seangle} and \ref{fig:seanglezoom}}, among the MSPs currently included in PTAs, the angularly closest MSP to the SMBHB candidate PKS 2131--021, which is the MSP J2145--0750, is such that its angular separation falls into a peak of the function $R\left(\theta\right)$. This makes it worthy of attention since in its TRs a periodic signature induced by the GWs emitted by the SMBHB candidate PKS 2131--021 may be present. For this reason, in Section \ref{sec:4}, the LS algorithm was used to search for such periodic signature. The periodograms plotted in Figures \ref{fig:narrow2} and \ref{fig:wide2} have been obtained from the complete datasets, provided by the NANOGrav collaboration \linebreak (see Figures \ref{fig:narrow1} and \ref{fig:wide1}), while the periodograms plotted in Figures \ref{fig:narrow4} and \ref{fig:wide4} have been obtained from the filtered datasets (see Figures \ref{fig:narrow3} and \ref{fig:wide3}), discussed in Section \ref{sec:3}. Interestingly, the LS algorithm finds in each dataset a periodicity associated with an angular frequency very close to the angular frequency of the GW emission expected for the SMBHB candidate PKS 2131--021. Although this result does not constitute proof of the SMBHB candidate PKS 2131--021 GW emission, it is certainly worthy of further investigation. A more detailed and rigorous analysis of the MSP J2145--0750 TRs, using the NANOGrav $12.5$ yr Data Set in conjunction with the IPTA Data Release 2 \cite{perera2019}, will be proposed in a future paper entirely dedicated to searching for GWs potentially emitted by the SMBHB candidate PKS 2131--021. There, it will also be discussed the use of other MSPs not too angularly far from the SMBHB candidate PKS 2131--021, as well as the possible improvements that will be brought on this kind of search by next-gen PTAs, such as Square-Kilometer Array (SKA) \cite{johnston2007,kramer2015,weltman2020,padmanabhan2022}.
%=================================================================

%=================================================================
\vspace{6pt} 
%=================================================================
%%%%%%%%%%%%%%%%%%%%%%%%%%%%%%%%%%%%%%%%%%
\authorcontributions{Conceptualization, M.M. and F.D.P.; Formal analysis, M.M. and F.D.P.; Writing---original draft, M.M.; Writing---review \& editing, M.M. and F.D.P.; Supervision, A.A.N. All authors have read and agreed to the published version of the manuscript.} %{For research articles with several authors, a short paragraph specifying their individual contributions must be provided. The following statements should be used ``Conceptualization, X.X. and Y.Y.; methodology, X.X.; software, X.X.; validation, X.X., Y.Y. and Z.Z.; formal analysis, X.X.; investigation, X.X.; resources, X.X.; data curation, X.X.; writing---original draft preparation, X.X.; writing---review and editing, X.X.; visualization, X.X.; supervision, X.X.; project administration, X.X.; funding acquisition, Y.Y. All authors have read and agreed to the published version of the manuscript.'', please turn to the \href{http://img.mdpi.org/data/contributor-role-instruction.pdf}{CRediT taxonomy} for the term explanation. Authorship must be limited to those who have contributed substantially to the work reported.}

\funding{This research received no external funding.}
\dataavailability{The data used in this paper were collected by the NANOGrav collaboration and are publicly available. %MDPI: please confirm if the reference citation is necessary in this section  ------> AUTHOR: It is not necessary, I remove it.
} 
\acknowledgments{We warmly acknowledge Andrea Possenti (INAF-OAC) for the enlightening discussions and comments on the paper. We acknowledge the support of the Theoretical Astroparticle Physics (TAsP) and Euclid projects of the Istituto Nazionale di Fisica Nucleare (INFN).}
\conflictsofinterest{The authors declare no conflict of interest.}
%%%%%%%%%%%%%%%%%%%%%%%%%%%%%%%%%%%%%%%%%%
%% Optional
\abbreviations{Abbreviations}{
The following abbreviations are used in this manuscript:\\

\noindent 
\begin{tabular}{@{}ll}
EPTA~~~~~~~~~~ & European Pulsar Timing Array\\
GW & Gravitational Wave\\
GWB & Gravitational Wave Background\\
INAF & Istituto Nazionale di Astrofisica\\
INFN & Istituto Nazionale di Fisica Nucleare\\
InPTA & Indian Pulsar Timing Array\\

\end{tabular}}

\abbreviations{}{%
\noindent
\begin{tabular}{@{}ll}
LIGO & Laser Interferometer Gravitational-Wave Observatory\\
LISA & Laser Interferometer Space Antenna\\
MDPI & Multidisciplinary Digital Publishing Institute\\
MSP & Millisecond Pulsar\\
NANOGrav & North American Nanohertz Observatory for Gravitational Waves\\
OAC & Osservatorio Astronomico di Cagliari\\
PPTA & Parkes Pulsar Timing Array\\
PTA & Pulsar Timing Array\\
SMBHB & Supermassive Black Hole Binary\\
SKA & Square Kilometre Array\\
TAsP & Theoretical Astroparticle Physics\\
ToA & Time of Arrival\\
TR & Timing Residual\\
\end{tabular}}
%=================================================================
%\end{paracol}
%%%%%%%%%%%%%%%%%%%%%%%%%%%%%%%%%%%%%%%%%%
\begin{adjustwidth}{-\extralength}{0cm}
\printendnotes[custom]
\reftitle{References}
%=================================================================
%\bibliography{bibliography.bib}

\end{adjustwidth}
\end{document}